\documentstyle[twoside,fleqn,espcrc2,rotate,epsf]{article}
\newcommand{\wh}{\widehat}
\newcommand{\IM}{\mbox{\rm Im}}
\newcommand{\nn}{\nonumber}

\newcommand{\as}{\alpha_{s}}
\newcommand{\ve}{\varepsilon}
\newcommand{\eqn}[1]{(\ref{#1})}
\newcommand{\mev}{\mbox{\rm MeV}}
\newcommand{\gev}{\mbox{\rm GeV}}
\newcommand{\MSb}{{\overline{MS}}}
\newcommand{\smvs}{\vbox{\vskip 8mm}}

\title{
\vspace{-1.6cm}
{\small\sf
\rightline{HD-THEP-97-51}
\rightline{hep-ph/9709484}}
\bigskip
\bigskip
{\bf The strange quark mass from scalar sum rules updated}}

\author{Matthias Jamin
\address{
{\em Institut f\"ur Theoretische Physik, Universit\"at Heidelberg}\\
{\em Philosophenweg 16, D-69120 Heidelberg, Germany}}}

\begin{document}

\begin{abstract}
The talk discusses preliminary results of an updated analysis of the
strange quark mass from the scalar current QCD sum rules \cite{jam:97}.
In particular the role of the scalar form factor which is a main ingredient
in the analysis is especially emphasised. The sources of the uncertainties
in the sum rule determination are briefly reviewed.
\end{abstract}

\maketitle


\section{Introduction}

A precise determination of the strange quark mass, being one of the
fundamental parameters in the Standard Model (SM), is of paramount interest
in several areas of present day particle phenomenology. Until today two
main methods have been employed to achieve this task. QCD sum rules
\cite{svz:79} have been applied to various channels containing strange
quantum numbers, in particular the scalar channel that will be the
subject of this talk \cite{jm:95,cdps:95,cps:96}, and more recently
also lattice QCD simulations have been used to extract the strange
quark mass \cite{all:97,ses:97}. 

The basic object which is investigated in the simplest version of
QCD sum rules is the two-point function $\Psi(q^2)$ of two hadronic
currents
\begin{equation}
\label{eq:1.1}
\Psi(q^2) \; \equiv \; i \int \! dx \, e^{iqx} \,
\big<0\vert \, T\{\,j(x)\,j(0)^\dagger\}\vert0\big>\,,
\end{equation}
where in our case $j(x)$ will be the divergence of the vector current,
\begin{equation}
\label{eq:1.2}
j(x) \; = \; \partial^\mu (\bar s\gamma_\mu u)(x) \; = \;
i\,(m_s-m_u)(\bar s\,u)(x) \,.
\end{equation}
$\Psi(q^2)$ is thus approximately given by $m_s^2$ times the two-point
function of the scalar current.

After taking two derivatives of $\Psi(q^2)$ with respect to $q^2$,
$\Psi''(q^2)$ vanishes for large $q^2$, and satisfies a dispersion
relation without subtractions,
\begin{equation}
\label{eq:1.3}
\Psi''(q^2) \; = \;
2 \int\limits_0^\infty \frac{\rho(s)}{(s-q^2-i\ve)^3}\,ds \,,
\end{equation}
where $\rho(s)$ is defined to be the spectral function corresponding
to $\Psi(s)$,
\begin{equation}
\label{eq:1.4}
\rho(s) \; \equiv \; \frac{1}{\pi}\,\IM\,\Psi(s+i\ve) \,.
\end{equation}

To suppress contributions in the dispersion integral coming from
higher excited states, it is further convenient to apply a Borel
transformation to eq.~\eqn{eq:1.3} \cite{svz:79}. The left-hand
side of the resulting equation is calculable in QCD, whereas under
the assumption of duality, the right-hand side  can be evaluated in
a hadron-based picture, thereby relating hadronic quantities
like masses and decay widths to the fundamental SM parameters.

Generally, however, from experiments the phenomenological spectral
function $\rho_{ph}(s)$ is only known from threshold up to some energy
$s_0$. Above this value, we shall use the perturbative expression
$\rho_{th}(s)$ also for the right-hand side. This is legitimate if
$s_0$ is large enough so that perturbation theory is applicable.
The central equation of our sum-rule analysis is then:
\begin{eqnarray}
\label{eq:1.7}
u^3\,\wh\Psi''_{th}(u) & = & \int\limits_0^{s_0} e^{-s/u}\rho_{ph}(s)
\,ds \nn \\
\smvs
& + & \int\limits_{s_0}^\infty e^{-s/u}\rho_{th}(s)\,ds \,.
\end{eqnarray}
In addition, in the analysis one can also use the first derivative of
this equation with respect to $u$ the so called ``first-moment sum rule''.

The main ingredients in these equations, namely the theoretical expression
for the two-point function as well as its phenomenological parameterisation,
will be discussed below.

\section{Theoretical two-point function}

In the framework of the operator product expansion the Borel
transformed two-point function $\wh\Psi(u)$ can be expanded in
inverse powers of the Borel variable $u$:
\begin{eqnarray}
\label{eq:2.1}
\wh\Psi(u) & \!\! = \!\! & (m_s-m_u)^2\,u\,\biggl\{\,\Psi_0(u)+
\frac{\Psi_2(u)}{u} \nn \\
\smvs
& & +\,\frac{\Psi_4(u)}{u^2}+\frac{\Psi_6(u)}{u^3}+\ldots\,\biggr\} \,.
\end{eqnarray}
The $\Psi_n$ contain operators of dimension $n$, and their remaining
$u$ dependence is only logarithmic.

The purely perturbative contribution $\Psi_0(u)$ is presently known up to
${\cal O}(\as^3)$, with the last term having been calculated
only very recently \cite{cps:96}. Numerically, the expansion reads
\begin{equation}
\label{eq:2.2}
\Psi_0(u) \, = \, \frac{3}{8\pi^2}\Big[1+1.53\as+2.23\as^2+
1.71\as^3\Big] .
\end{equation}
In this expression the strong coupling constant $\as(u)$ should be evaluated
at the scale $u$.
Therefore, even for $\as(1\,\gev)\approx 0.5$ the last term is roughly
20\% and the perturbative expansion displays a reasonable convergence.
Because the two-point function scales as $m_s^2$, the resulting
uncertainty for $m_s$ from higher orders is at most 10\%. In practice
it is somewhat smaller since the average scale at which the sum rule is
evaluated lies around $1.5\,\gev$.

The next term in the operator product expansion $\Psi_2(u)/u$ only receives
contributions proportional to the quark masses squared. Already at a
scale of $u=1\,\gev^2$ its size is approximately 2\%, decreasing like $1/u$
for higher scales. Although it has been included in the phenomenological
analysis, for the error estimates on the strange quark mass it can be
safely neglected.

The same holds true for the dimension-four operators. Here there are
contributions from the quark and gluon condensates as well as
explicit mass corrections $\sim\!\! m^4$. Again, at a scale of $u=1\,\gev^2$
the size of $\Psi_4(u)$ is below 1\% of the full two-point function,
hence being negligible for the $m_s$ analysis. Nevertheless, the
$\Psi_4$ and in addition the $\Psi_6$ contributions have been
included for the numerical investigations.

\section{Hadronic spectral function}

Generally, all intermediate states with the correct quantum numbers
contribute to the hadronic spectral function. In the case of the scalar
two-point function the lowest lying state is the $K\pi$-system in an
$s$-wave isospin $1/2$ state. The contribution of this intermediate
state yields the inequality \cite{jm:95}
\begin{equation}
\label{eq:3.1}
\rho(s) \, \ge \, \frac{3\,\theta(s-s_+)}{32\pi^2\, s}\,
\sqrt{(s-s_+)(s-s_-)} \,\big|d(s)\big|^2
\end{equation}
where
\begin{equation}
\label{eq:3.2}
s_+ \; = \; (M_{\rm K} +M_{\rm \pi})^2 \;,
\quad
s_- \; = \; (\rm M_{\rm K}-M_{\rm \pi})^2 \;,
\end{equation}
and $d(s)$ is the strangeness-changing scalar form factor
\begin{equation}
\label{eq:3.3}
d(s) \; = \; -\,i\sqrt{2}\,\langle\pi^0K^-|\partial^\mu(\bar s\gamma_\mu u)(0)
|0\rangle
\end{equation}
also appearing in $K_{l3}$ decays. The scalar form factor $d(s)$ admits
an Omn\`es representation which can be found in \cite{jm:95} and
depends on the $K\pi$ $s$-wave, $I=1/2$ phase shift $\delta_0^{1/2}$.
Similarly to an analysis of the pion form factor \cite{gp:97} the
Omn\`es representation can be improved by using knowledge on effective
hadronic theories and the $1/N_c$ expansion thereby fixing a polynomial
ambiguity which exists in the Omn\`es representation \cite{jam:97}:
\begin{eqnarray}
\label{eq:3.4}
d(s) & \!\!=\!\! & \frac{d(0)M_R^2}{(M_R^2-s-iM_R\Gamma_R(s))}\cdot \nn \\
\smvs
& & \exp\Biggl\{\,\frac{s}{\pi}\int\limits_{s_+}^\infty
\frac{\delta_{0,\,bg}^{1/2}(t)}{t(t-s-i\ve)}\,dt\,\Biggr\} \,,
\end{eqnarray}
with
\begin{equation}
\label{eq:3.5}
d(0) \; = \; 0.977\cdot(M_{\rm K}^2-M_{\rm \pi}^2) \,,
\end{equation}
$M_R$ and $\Gamma_R(s)$ are the mass and energy dependent width of
the lowest lying resonance, the $K_0^*(1430)$ in this case, and
$\delta_{0,\,bg}^{1/2}$ is the background contribution to the $s$-wave phase
shift which can be extracted from experimental data on $K\pi$ scattering.

Further details on this representation of the scalar form factor and a
discussion of the deficiencies of the representation used in ref.~\cite{jm:95}
can be found in \cite{jam:97}.

\begin{figure}[thb]
\vspace{-18mm}
\centerline{
\rotate[r]{
\epsfxsize=2.8in
\epsffile{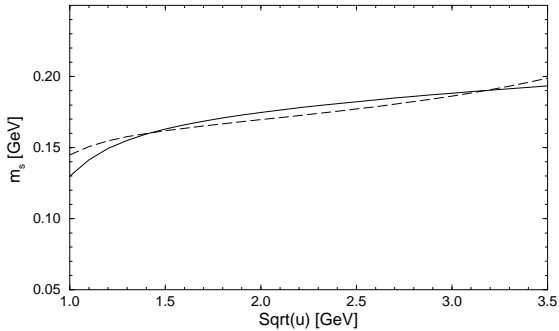} }}
\vspace{-18mm}
\caption[]{The strange quark mass obtained from the zeroth and first
moment sum rules. \label{fig:1}}
\end{figure}

\section{Phenomenological analysis}

Evaluating the sum rule of eq.~\eqn{eq:1.7} and the corresponding first
moment sum rule with the theoretical two-point function of section~2
and the hadronic spectral function of section~3, the resulting values
for the running strange quark mass $m_s(1\,\gev)$ in the $\MSb$ scheme
as a function of $\sqrt{u}$ are displayed in figure~1 (solid and dashed
lines respectively). The continuum threshold $s_0$ has been determined
to be approximately $s_0=3.4\,\gev^2$ by requiring duality, namely equality
of the strange masses obtained from the zeroth and first moment sum rules.
A value of this size is also expected from the fact that in this region
the second resonance, the $K_0^*(1950)$, which has not been included in
the analysis, is found.

On the other hand, in an interval for $u$ where we expect the sum rules
to be valid, the sum rules should be stable and the extracted strange
quark mass should be independent of $u$. However, the stability of the
curves shown in figure~1 is not very good. This is due to the fact that
the region where duality holds overlaps with the region of the second
resonance which should thus be included. The contribution of multi-particle
intermediate states like $K\pi\pi\pi$ is of higher order in the chiral
expansion and should be suppressed.

Estimating the error from the variation of the strange mass in
the range $1\,\gev^2<u<9\,\gev^2$, as a preliminary result from our
analysis we find $m_s(1\,\gev)=160\pm30\,\mev$. This indicates that
the dominant error in the determination of the strange quark mass
stems from the parameterisation of the hadronic spectral function.
Compared to this error the uncertainties from higher order $\as$
corrections are small. A detailed discussion of the phenomenological
spectral function as well as the inclusion of the second resonance
can be found in ref.~\cite{jam:97}.

\bigskip \noindent
{\bf Acknowledgements}
The author would like to thank S. Narison for the invitation to
this very pleasant and interesting conference.


\bigskip \noindent
{\bf Discussions}

\medskip
\noindent {\bf P. Raczka}, Durham

\noindent
{\em What is the theoretical uncertainty in the final result corresponding
to the two-parameter freedom in the choice of the renormalization scheme
for the next-next-to-leading order perturbative QCD correction?}

\medskip
\noindent {\bf M. Jamin}

\noindent
{\em The scale uncertainty has been estimated to be about 5\% by varying
the renormalization scale. An error for the dependence of the quark mass
on the renormalization scheme has not been included in the analysis.

\end{document}